# Monte Carlo Simulations of Factors Influencing Seasonal Variation of Multiple Muon Events

Ellen Guan
Illinois Math and Science Academy and Argonne National Laboratory

Dr. Maury Goodman
Argonne National Laboratory

**Abstract**
　　In a 2015 study, particle physicists reported the first observed seasonal variation of multiple muon events in the MINOS Far and Near Detectors, where multiple muon events created by cosmic rays were observed to be more numerous in the winter than in the summer. This goes against researchers' initial hypothesis, because it has long been measured that single muon rates are higher in the summer. Using Python, I simulated three potential factors that could affect this: muon decay, altitude, and the fate of the leading pion from the first interaction. My data gives support to the conclusion that the seasonal variation is due to the geometry of the underground particle detectors and related to the higher altitude of the atmosphere in the summer.

**Introduction**
　　Cosmic rays are particles in our galaxy that have been accelerated close to the speed of light and can be found traveling in any direction. Some cosmic rays reach the Earth's atmosphere, where they can collide with a nucleus of an atom found in the atmosphere. This collision can create the particles electrons, neutrons, pions, and kaons. Pions and kaons, in particular, can decay into particles known as muons, and these particles can penetrate through Earth, where they can be detected. However, the more dense the location in the atmosphere where the pions and kaons are formed, the more likely it is for the particles to interact and therefore lose enough of the energy to prevent them from creating detectable muons. Since the atmosphere expands when it warms in the summer, more muons are expected to be detected in the summer (Adamson et al, 2015). This has been widely observed among multiple underground particle detectors.

In a 2015 study, *Observation of seasonal variation of atmospheric multiple-muon events in the MINOS Near and Far Detectors*, scientists reported the first observed seasonal variation of multiple muon events in the MINOS Far and Near Detectors, where multiple muon events created by cosmic rays were observed to be more numerous in the winter than in the summer. This goes against the expectation, since, according to the ideal gas law, the atmosphere is less dense in the summer than in the winter, therefore the pions and kaons would be less likely to interact in the summer before they decay (Adamson et al, 2015).

Some scientists on the NOvA experiment currently believe that this observation is due to the geometry of the underground particle detectors (Acero et al, 2019), rather than two other proposed explanations, which are muon decay and the starting proton. To test this, over a course of a few months at Argonne National Laboratory, I use Python to simulate the amount of winter versus summer multiple-muon events when impinging the top of the NOvA Near Detector depending on these three different effects. The physical calculations involve the geometrical factors of the detector, including the number of multiple-muon events versus the separation length between the muons at the detector and number of multiple-muon events versus altitude in the atmosphere of when the muons are first created, since the higher up the pions and kaons are created, the more likely it is for them to be able to decay into a muon that could reach Earth's surface. Both of these physical calculations can help support the geometry effect hypothesis. I also calculate if muons are more likely to decay in the summer than the winter before ever reaching the detector, which may support the muon decay hypothesis. Lastly, I simulate if there are more multiple muon events detected when the first proton decays rather than interacts, which may support the starting proton hypothesis.

**The Geometry Effect**

In the 2015 study observing the first seasonal variation of multiple muon events, the scientists proposed four different explanations to the phenomenon, one being the geometry effect (Adamson et al, 2015). The altitude of the first point of muon creation is related to the absolute temperature, where a 2% change in temperature will cause a 2% change in altitude. Essentially, since when the atmosphere expands in the summer, the air has nowhere to go except up. Thus, muons created at a certain point in the winter will be generated at a higher point in the summer due to the less dense atmosphere, as shown in Figure 1. Since underground particle detectors have a finite size, a more spread out separation distance between generated muons could mean that while one muon may be detected by the detector, the other muon may miss the underground detector entirely.

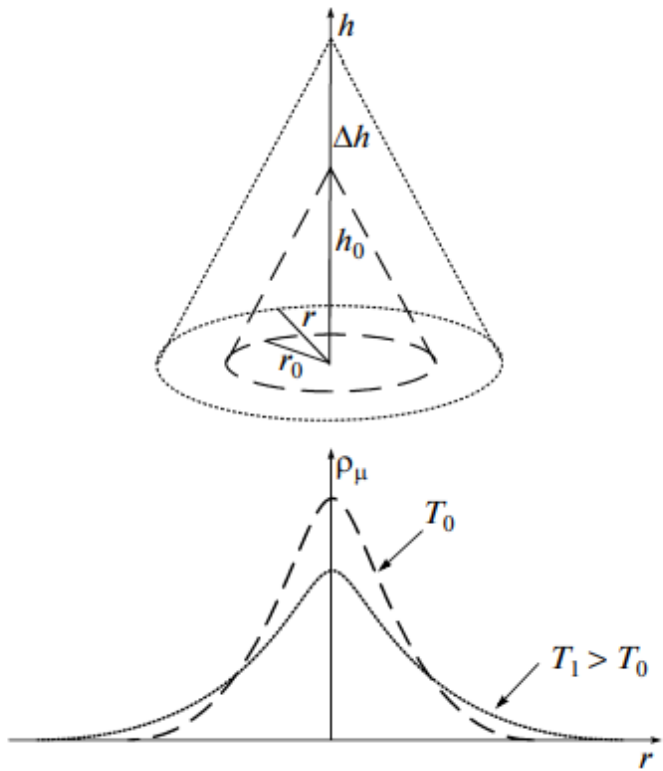

**Figure 1**: Illustration of muon distribution as height and temperature increases (Tolkacheva, N. V., Bogdanov, A. G., Dmitrieva, A. N., Kokoulin, R. P., & Shutenko, V. V., 2011)

**Effect of Muon Decay**

If muons are created higher up in the atmosphere, it takes more time for them to reach the surface to be detected. One hypothesis that can explain the winter multiple muon events max is the fact that since muons are created 2% higher in the atmosphere during the summer than the winter, more muons may decay before reaching the underground particle detector during the summer.

To test this, I used the equation shown below to calculate the number of relativistic muons left over at fifteen kilometers as measured from sea level depending on the muon's initial energy.

$$N = N_0 \, e^{(-t/\gamma\tau)} = e^{(-ct/\gamma c\tau)}$$

N is the number of muons left that haven't decayed
t is time
$\tau$ is 2.2 10^-6 s
c is the speed of light
$\gamma = E_\mu/m_\mu = E_\mu/106 \text{ MeV} = 472$
ct (winter) = 15 km
ct (summer, due to a 2% increase) = 15.3 km

Using these values, the values in Figure 2 and 3 can be calculated:

| Energy of Muon (GeV) | Percent of Muons that Haven't Decayed in the Winter | Percent of Muons that Haven't Decayed in the Summer |
|---|---|---|
| 5 | 62.05% | 61.46% |
| 50 | 95.34% | 95.25% |
| 500 | 99.52% | 99.51% |

**Figure 2**: Table of the percent of muons that don't decay depending on the starting energy of a muon

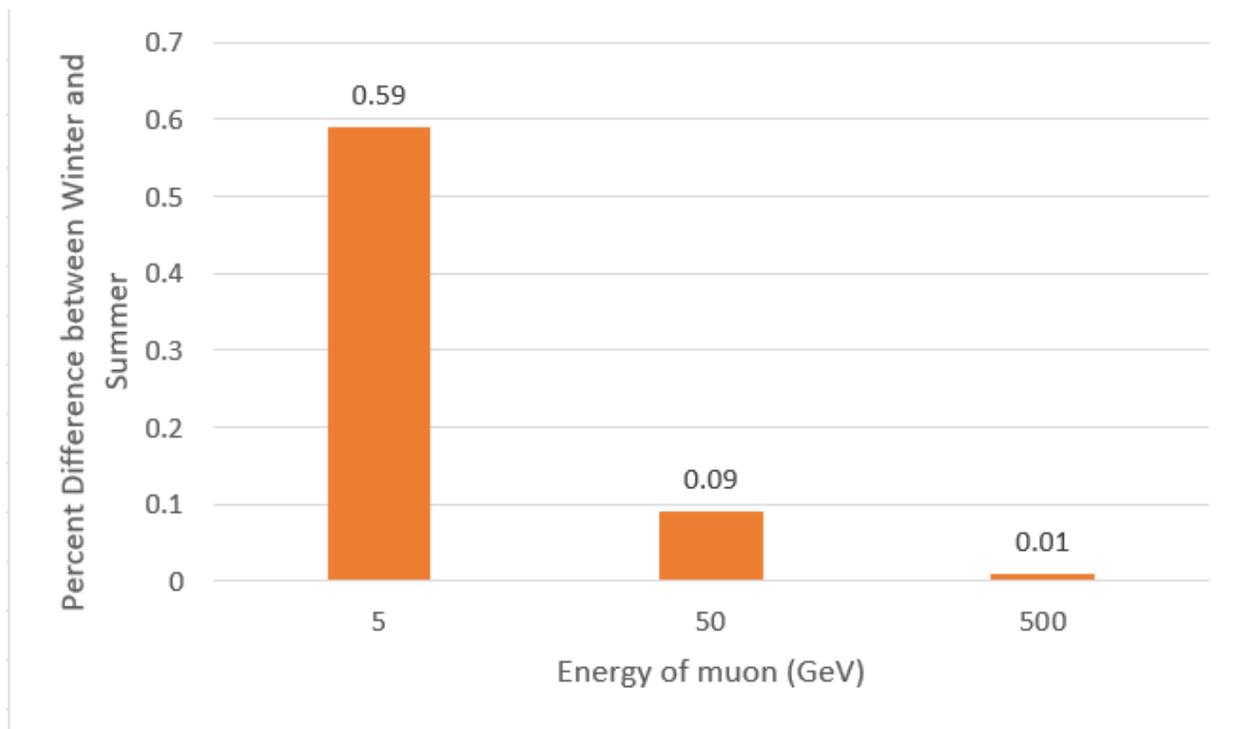

**Figure 3:** Percent of muons that don't decay depending on the starting energy of a muon

      As shown in Figures 2 and 3, between summer and winter at the critical pion energy of 50 GeV, there is only a .1% difference, which is too small to account for a 2% seasonal variation. The NOvA Near Detector is able to detect any muons with initial energies greater than 50 GeV, and as shown by Figure 2, increasing the energy of the muon will only decrease the summer and winter difference.

**Multi-Muon Separation Distance**
      Due to the finite size of the underground particle detectors, when the separation distance of two muons observed by the detectors increases, the amount of total multiple muon event pairs decreases. Using the data for the amount of multiple muon pairs observed as a function of the muon separation distance at an altitude of 15 kilometers collected by the NOvA Near Detector, I calculated the difference in separation distance between winter multiple muon pairs and summer pairs. Each separation distance determined by a winter multiple muon pair will be 2% smaller than the separation distance of a summer multiple muon pair. The data I used includes 100 bins of the square of muon separation and was calculated by a private reference using CORSIKA, a pre-packaged sophisticated Monte Carlo physics computer software that simulates air showers as created by high energy cosmic rays. The simulation used the dimensions of the NOvA Near Detector, and the histogram of such data is reproduced in Figure 4. The muons come straight down (zenith angle of 0 degrees) onto the detector.

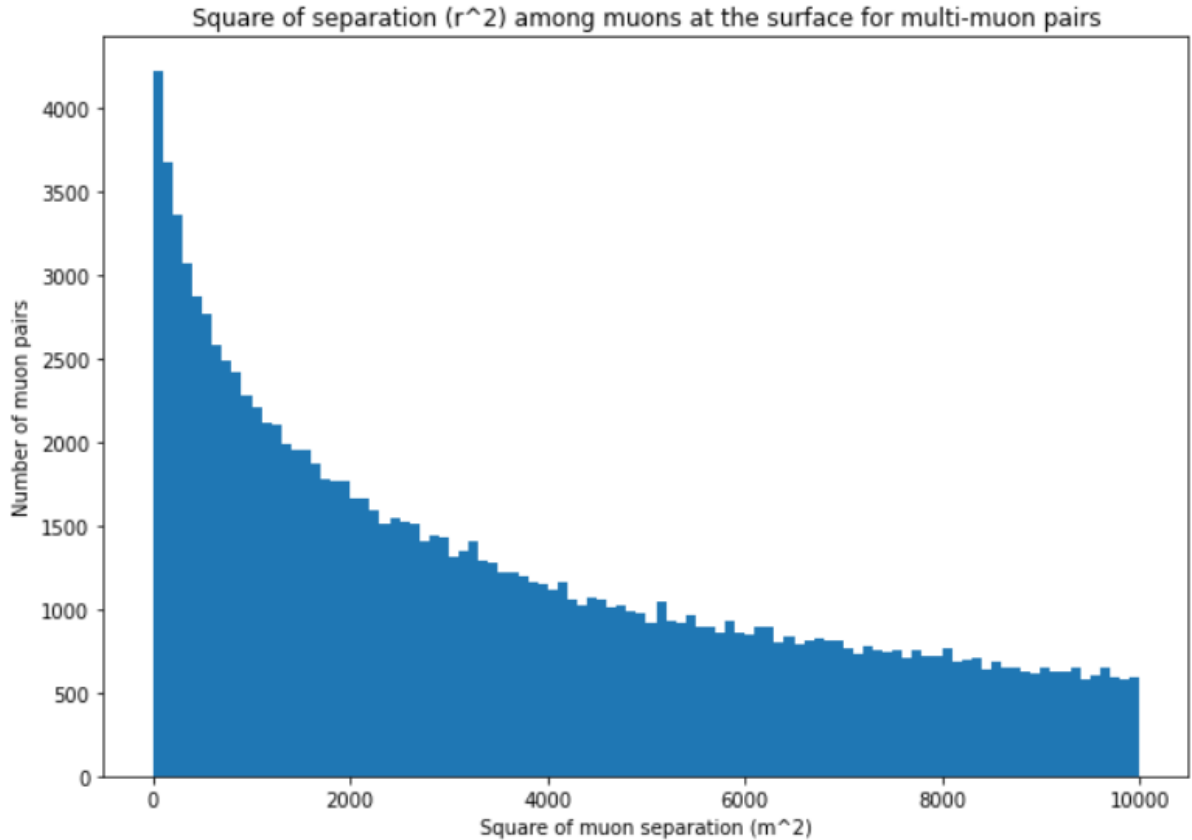

**Figure 4:** Histogram of the square of separation in meters squared among muons at the surface of the NOvA Near Detector for multi-muon pairs

      To calculate the separation distance percent difference, I used Python to create a simulation that chooses a random square separation distance and a random number of muon pairs included in the data. Then, the simulation chooses a random theta angle and a random point (x-direction and y-direction) that would fit onto the surface of the NOvA Near Detector, which is around 4 meters widthwise by 15 meters lengthwise. This random point represents a single muon detected on the surface of the particle detector. Next, the simulation checks if a second muon could fit on the detector surface and therefore be detected from the previously chosen distance away from the first muon. If the second muon can be detected, this counts as one winter multiple muon event observed. The simulation goes on to check if, using the same random point but instead with an increased separation distance by 2% for the second muon, the resulting muon pairs would both be able to be detected by the particle detector. If this is true, this counts as one summer multiple event observed. For the simplicity of this simulation, I only took into account muons hitting and therefore being detected by the top 2D surface of the particle detector, and I rounded the surface to be 4 meters by 15 meters. I ran this simulation for 5,000,000 trials, and calculated a percent difference from the winter multiple muon events and the summer winter multiple muon events to be 3.8% with a statistical error of ± 0.1%, as shown in Figure 5.

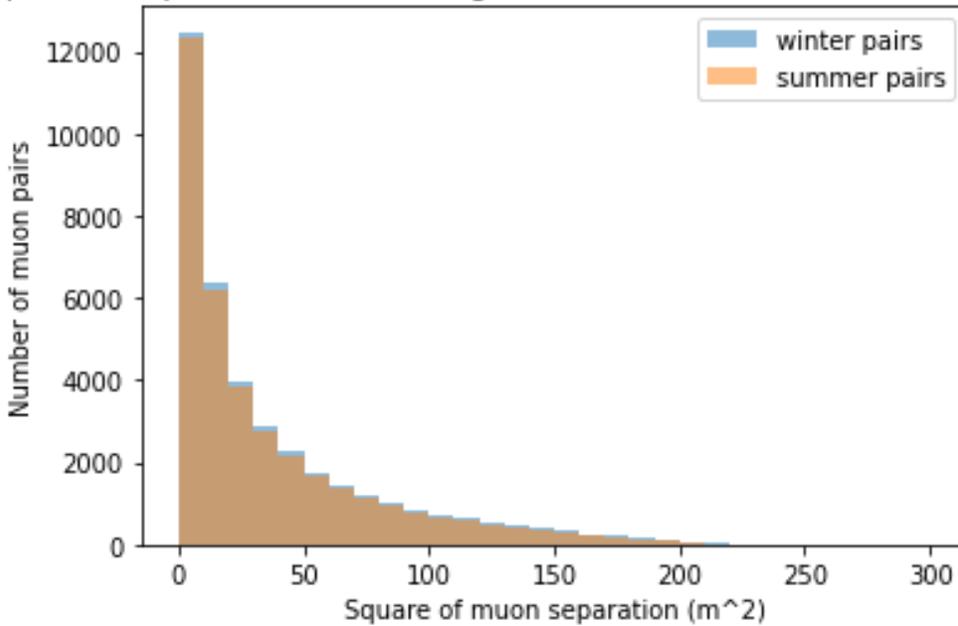

**Figure 5:** Histogram of the number of winter multi-muon pairs versus summer multi-muon pairs as a function of the square muon separation

**Altitude**

Using the same data from before except with 1,000 bins of the square of muon separation distance data, I used the same simulation to calculate the number of total winter multiple muon pairs and total summer multiple muon pairs but with a variety of different altitudes. Varying the altitude in which a muon is created increases the muon pair separation distance proportionally. I ran the simulation for 10,000,000 trials, and my results are shown in the table in Figure 6 and 7. It is seen that there is a clear altitude dependence to the seasonal variation, and that at the altitude of 15 km where many muons are created, the magnitude of effect matches what was seen in the MINOS and NOvA data.

| Altitude | Separation percentage | (Winter pairs, Summer pairs) | (W-S)/(W+S) Percent difference |
|---|---|---|---|
| 60 km | 8% | (577438, 503426) | 6.80% |
| 30 km | 4% | (576733, 537156) | 3.60% |
| 15 km | 2% | (577954, 557511) | 1.80% |
| 7.5 km | 1% | (577536, 567298) | 0.90% |
| 3.75 km | 0.50% | (577910, 572838) | 0.40% |

**Figure 6:** Table of the number of winter and summer multi-muon pairs as a function of altitude and the calculated percent difference

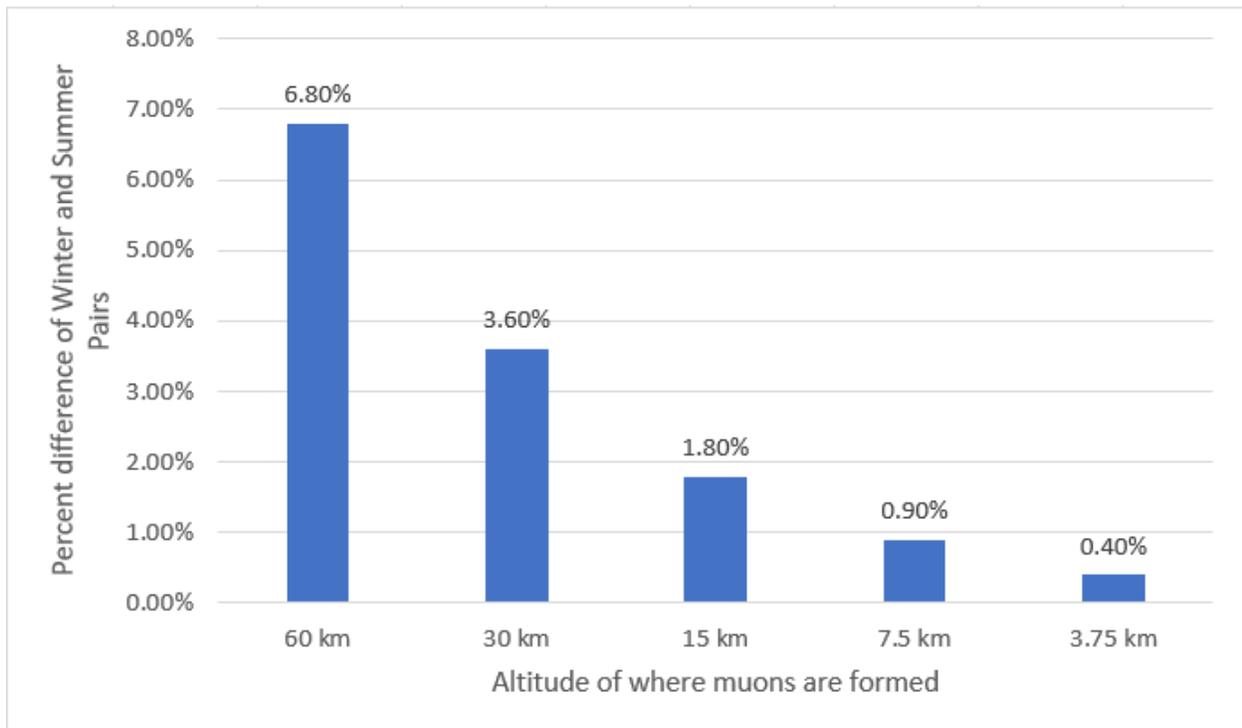

**Figure 7**: Calculated percent difference of winter and summer pairs versus altitude of where muons are formed

**Energy of the Starting Proton**
      The data I used for my simulations involving the multiple muon event separation distance and altitude is experimental, so it may include muons of a variety of different energies that have been detected. Thus, it is valuable to systematically study the effect of the amount of muons created when the starting proton (from a cosmic ray and interacts with the nucleus of an atom in the atmosphere which leads to the creation of muons) has different energies, since every interaction of particles decreases their energy. This is also important because since single muon events peak in the summer where the first pion decays, this means that single muon events have a minimum in the winter where the first pion interacts. Thus, there may be a winter maximum for multiple muons because when the first pion interacts, it can give more opportunity for more muons to be detected, albeit these muons have less energy than if the first pion decays. This is

one of the hypotheses that can explain the winter versus summer multiple muon events difference.

To test this, I created a simulation using Python to calculate the average number of pions that decay into muons when the first pion (that is created after the first proton interacts) decays and the number that decay into muons when the first pion interacts. In this simulation, kaons are not taken into account for simplicity. The simulation made the following assumptions:

When a proton from a cosmic ray enters the atmosphere and interacts with the air, it can decay into one of four different cases (each with 25% probability):

1. a positive pion, a neutron, and air
2. a positive pion, a negative pion, a proton, and air
3. a neutral pion, a proton, and air
4. a positive pion, a neutral pion, a neutron, and air

In these cases, the proton or neutron will get 50% of the initial proton's energy, and the other particles (except for air) will receive the remaining energy split evenly. My simulation ignores what happens with air and neutral pions, since they do not interact or decay into anything relevant to the muon multiplicity. The resulting positive or negative pions, neutrons, and protons each travel down to the next interaction point where they each can either decay or interact. For neutrons, the four cases (with each case having a 25% probability) are:

1. negative pion, proton, and air
2. negative pion, neutral pion, proton, and air
3. neutral pion, neutron, and air
4. positive pion, negative pion, neutron, and air

Like the proton, the resulting proton or neutron will get 50% of the initial neutron's energy, and the other particles (except for air) will receive the remaining energy split evenly. For positive pions, the four cases (each with 25% probability) are:

1. positive pion, neutral pion, and proton
2. positive pion, positive pion, and neutron
3. positive pion, negative pion, neutral pion, and proton
4. positive pion, positive pion, neutral pion, and neutron

For negative pions, the four cases (each with 25% probability) are:

1. a negative pion, neutral pion, and proton
2. negative pion, positive pion, and neutron
3. negative pion, positive pion, negative pion, proton
4. negative pion, positive pion, neutral pion, and neutron

For positive/negative pions, one resulting pion will get 50% of the initial pion's energy, and the remaining particles will receive the remaining energy split evenly. In terms of the probability of a pion interacting with another particle at an interaction point instead of decaying into a muon, it

follows this formula, which is the general formula for a lifetime of a particle:

$$1 - e^{-\frac{ct}{\gamma c\tau}}$$

where $\tau = 2.6 \times 10^{-6}$ s which is the lifetime of the pion
$c\tau = 7.8$ m
$\gamma$ = (Energy of the pion)/140 MeV
ct = 25 km / 12.8 ≈ 2 km which is the average distance between interactions in the upper atmosphere

In this simulation, a lot of specifics are greatly simplified and rounded. The atmosphere density is taken to be consistent, and interaction points are taken to be once every ~2 km. Since the critical pion energy is 135 GeV, where the probability of a pion decaying or interacting is the same, the above formula was greatly simplified to account for the decay probability where critical seasonal variation may occur:

$$1 - e^{-\frac{665}{\gamma}}$$

At every interaction point, if the simulation has created a positive or negative pion, the simulation chooses a random number between 0 and 1, and if the number if less than the above calculated decay probability, the current pion decays into a muon with 40% of the pion's current energy (the rest of the energy goes into neutrinos). The simulation counts any muons with energies of over 50 GeV, since these muons have enough energy to be detected by the NOvA underground particle detector at Fermilab. I ran this simulation for various starting energies of the first proton and collected the data shown in Figure 8 and 9.

| Energy of starting proton (GeV) | Average Number of Pions that Decay When the First Pion Decays | Average Number of Pions that Decay When the First Pion Interacts |
|---|---|---|
| 300 | 1.0 (65 trials) | 0.2 (35 trials) |
| 500 | 1.0 (56 trials) | 0.14 (44 trials) |
| 700 | 1.14 (44 trials) | 0.52 (56 trials) |
| 1000 | 1.17 (29 trials) | 0.49 (71 trials) |
| **3000** | 1.71 (14 trials) | 1.72 (86 trials) |
| **10000** | 2.6 (5 trials) | 4.81 (95 trials) |

**Figure 8:** Table of the average number of pions that decay into muons when the first pion decays versus interacts

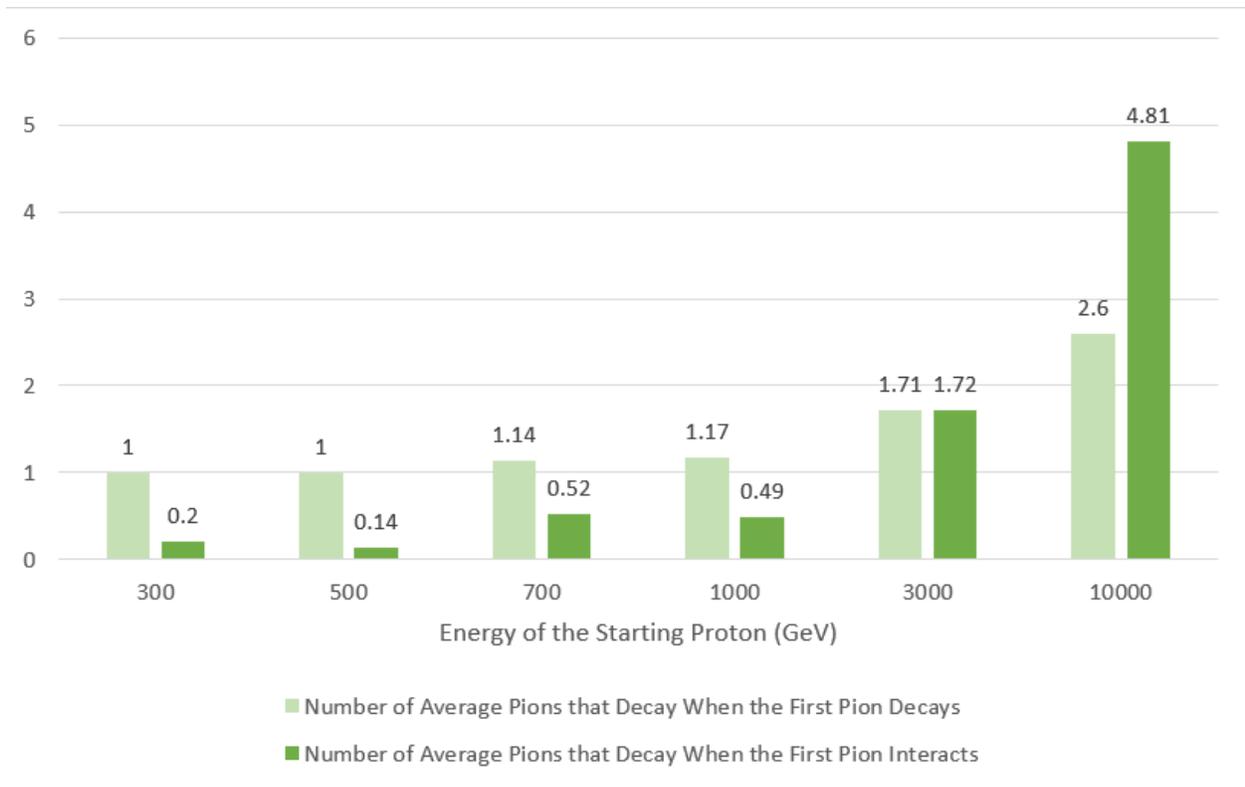

**Figure 9:** Number of average pions that decay when the first pion decays and interacts versus energy of the starting proton

**Discussion**

    My results support my initial and researchers' initial hypothesis that observed multiple muon events peak during the winter rather than the summer because of geometrical factors of the underground particle detectors and provide evidence against the muon decay and the starting proton hypothesis. As shown in Figures 6 and 7, there are more winter muon pairs calculated in the simulations regardless of altitude or multi-muon separation distance. As shown in Figures 8 and 9, there are still more muons that can be detected when the first pion decays rather than interacts, and only with much higher energies of the leading proton shows more muons being detected when the first pion interacts. However, the lower flux of higher energy proton cosmic rays will result in a much smaller muon flux in general, so if there were more muons when the first pion decays when the starting proton has a small energy, it would be able to help explain the

discrepancy between the seasonal variation of single and multiple muon events. Because this isn't the case, my results give support against the starting proton hypothesis. As shown in Figures 2 and 3, there are more muons that have not decayed in the winter than in the summer, however the percent difference does not account for the 2% difference observed, thus this data gives support against the muon decay hypothesis.

There are multiple limitations to my findings, however. In all of my Monte Carlo simulations, many assumptions were taken to simplify each particular situation. For example, I calculated the number of winter multiple muon events in contrast to the number of summer multiple muon events versus altitude and separation distance with only the surface of the underground particle detector instead of the 3D one. I also assumed the density of Earth's atmosphere increases linearly instead of exponentially in my starting proton simulation. While these assumptions may be made to produce a general result, these simplified simulations may not be able to predict completely accurate results taking into consideration the greater complexity of Earth's actual conditions.

Although my findings support the geometry effect solution, this does not mean that the geometry effect is indeed the solution to the multiple muon event winter maximum phenomena. There are still questions left unanswered, including the fact that while my data shows that there are more multiple muon pairs during the winter because there is a larger muon pair separation distance, CORSIKA sees a larger muon pair separation distance in the summer than the winter. CORSIKA also sees that when iron is the starting proton, there is a winter max of muon events that is detected by the infinite detector, even if when iron interacts, it splits into fifty particles. Lastly, CORSIKA sees a different single muon event max during the winter and the summer, which cannot be explained by the geometry effect since the geometry doesn't apply with only single muon detection. Further research can be done to support the geometry effect solution, including considering the effects of the earth's magnetic field, zenith angle, and multiple muon scattering on the seasonal multiple muon events distribution. More data on this phenomenon can also be collected in the future Deep Underground Neutrino Experiment because the detectors are much deeper in the ground and can detect muons with much higher energies.